\documentclass[twocolumn,prl,aps,showpacs,floatfix]{revtex4-1}
\usepackage{graphicx}

\usepackage[latin1]{inputenc}
\usepackage{amsmath}
\usepackage{amssymb}
\usepackage{amscd}
\usepackage{bbm}
\usepackage[all,cmtip]{xy}
\usepackage{color}

\newcommand{\mathd}{\mathrm{d}}
\newcommand{\mathe}{\mathrm{e}}

\newcommand{\tmop}[1]{\ensuremath{{#1}}}
\newcommand{\tmtextbf}[1]{{\bfseries{#1}}}
\newcommand{\tmtextit}[1]{{\itshape{#1}}}

\begin{document}

\title[]{Non-Markovian master equations from piecewise dynamics}

\author{Bassano Vacchini}

\affiliation{Dipartimento di Fisica, Universit{\`a} degli Studi di
Milano, Via Celoria 16, I-20133 Milan, Italy\\
INFN, Sezione di Milano, Via Celoria 16, I-20133
Milan, Italy}

\email{bassano.vacchini@mi.infn.it}

\begin{abstract}
We construct a large class of completely positive and trace preserving
non-Markovian dynamical maps for an open quantum system. These maps arise
from a piecewise dynamics characterized by a continuous time evolution
interrupted by jumps, randomly distributed in time and described by a quantum
channel. The state of the open system is shown to obey a closed evolution
equation, given by a master equation with a memory kernel and a inhomogeneous
term. The non-Markovianity of the obtained dynamics is explicitly assessed
studying the behavior of the distinguishability of two different initial
system's states with elapsing time.
\end{abstract}


\maketitle

Open quantum systems naturally arise in quantum mechanics due to lack of
isolation, and one of the basic difficulties in the field is the derivation of
closed irreversible evolution equations for the system only, taking into account the
interaction with the environment {\cite{Breuer2007,Weiss2008,Holevo2001}}. In particular an
open issue is the characterization and study of memory effects described by
these irreversible dynamics. An important class
of dynamical evolutions is given by
quantum dynamical semigroups, which by construction ensure complete
positivity (CP) and have a number of attracting physical and
mathematical features. The semigroup property ensures the existence of a closed evolution equation,
known as master equation, whose general expression has been determined
in the 70's just thanks to the requirement of CP
{\cite{Gorini1976a,Lindblad1976a}}.
The operators appearing in the the master equation
can be easily linked to the microscopic events which characterize the
dynamics. Moreover the exact solution can be expressed in terms of a
Dyson expansion, which allows for a natural reading in terms of a piecewise
dynamics consisting of a relaxing evolution interrupted by jumps.

In this Letter we show how a similar construction can be exploited
to obtain a large class of non-Markovian completely positive trace preserving
(CPT) maps, still admitting closed evolution equations. The building blocks of
this construction are a collection of time dependent
maps, together with a waiting time distribution describing the random
occurrence in time of interaction events described by a quantum channel. The
operational construction provides a direct physical reading of the different
contributions to the dynamics. The resulting master equations exhibit an
integral kernel which warrants CP of the solution, one of the
crucial difficulties in looking for extensions of the Lindblad result
{\cite{Barnett2001a,Budini2004a,Breuer2008a,Breuer2009a}}.

\paragraph*{Master equations.}

For a semigroup we have $\rho (t) = \Phi (t) \rho$, where
the time evolution operator obeys the master equation $\mathd
\Phi (t) / \mathd t = \mathcal{L} \Phi (t)$ and satisfies
\begin{displaymath}
  \Phi ( t_1 + t_2 ) = \Phi ( t_2 ) \Phi ( t_1
  ), \hspace{2em} \forall t_1, t_2 \geqslant 0.
\end{displaymath}
Introducing a self-adjoint operator $H$ and operators $L_k$ that can
be associated to microscopic interaction events, e.g. the exchange of an
excitation between system and bath, the operator
$\mathcal{L}$ called the generator takes the form {\cite{Gorini1976a,Lindblad1976a}}
$\mathcal{L} \rho = R \rho + \rho R^{\dag} + \mathcal{J} \rho$, 
where $R = - i  H -  (1/2) \sum_k
L_k^{\dag} L_k$, and the CP superoperator $\mathcal{J}$ reads
\begin{displaymath}
  \mathcal{J} \rho = \sum_k L_k \rho L_k^{\dag} .
\end{displaymath}
Introducing further the superoperator $\mathcal{R} (t)$, which
gives the semigroup obtained exponentiating the operator $R$
\begin{displaymath}
  \mathcal{R} (t) \rho = \mathe^{tR} \rho \mathe^{tR^{\dag}},
  \nonumber
\end{displaymath}
the exact evolution can be written as the Dyson series
\begin{eqnarray}
  \Phi (t) \rho & = & \mathcal{R} (t) \rho + \sum_{n
  = 1}^{\infty} \int^t_0 \mathd t_n \ldots \int^{t_2}_0 \mathd t_1 
  \label{eq:dyson}\\
  &  & \hspace{2em} \times \mathcal{R} ( t - t_n ) \mathcal{J}
  \mathcal{R} ( t_n - t_{n - 1} ) \ldots \mathcal{J} \mathcal{R}
  ( t_1 ) \rho . \nonumber
\end{eqnarray}
Here $\rho$ denotes the reduced system state taken as initial condition, and
the result follows from the Schwinger formula {\cite{Karplus1973a}} granting
in particular trace preservation. This solution can be naturally described as
a sequence of jumps, corresponding to transformations induced by the
CP map $\mathcal{J}$, distributed over an underlying relaxing
evolution given by the semigroup $\mathcal{R} (t)$. This kind of dynamics is universally accepted as Markovian. Indeed the fact that the state of the system at a
time $t_1 + t_2$ only depends on its state at a previous time $t_1$
expresses a feature that is naturally associated to lack of memory and
therefore to Markovianity (M). In this sense also a collection of two time evolution maps
$\Phi ( t + \tau, t )$ obeying the composition law
\begin{displaymath}
  \Phi ( t_1 + t_2, 0 ) = \Phi ( t_1 + t_2, t_1 )
  \Phi ( t_1, 0 ), \hspace{2em} \forall t_1, t_2 \geqslant 0
  \nonumber
\end{displaymath}
where each map is CPT, embodies the same idea of independence from the states
at previous times, and is therefore taken as a natural criterion to assess or
define M, known as divisibility {\cite{Rivas2010a,Chruscinski-xxx}}. Most
recently a novel idea has been put forward to characterize M, neither basing
on a representation of the dynamics, nor on the notion of memory as dependence
on the previous states of the system, but rather on the notion of
distinguishability of system's states, and on its behavior in the course of
the dynamics, which calls for an involvement of the environment and of
correlations {\cite{Breuer2009b,Breuer2012a}}. It turns out that this
criterion is satisfied by a dynamics characterized by divisibility, but is in
general less restrictive
{\cite{Laine2010a,Mazzola2010a,Haikka2011a,Vacchini2011a}}.

\paragraph*{Derivation from piecewise dynamics.}

We now build on these known results to construct a much wider class of time
evolutions, which admit a natural reading in terms of a piecewise dynamics,
with microscopic interaction events embedded in a continuous time dynamics.
These dynamics obey closed evolution equations expressed by means of a master
equation, possibly admitting an inhomogeneous contribution, which keeps track
of the initial condition. As a starting point we
consider Eq.~(\ref{eq:dyson}), replacing the semigroup
$\mathcal{R} (t)$ with a collection of time dependent CPT maps
$\mathcal{F} (t)$, which describe the time evolution between
jumps.
The events taking place over the background of the continuous time evolution
are described by a CPT map $\mathcal{E}$, namely a quantum channel, and their
distribution in time is characterized by an arbitrary waiting time
distribution, so that the number of events in time realizes a renewal process.
In terms of these basic building blocks one has, given an initial
state $\rho$, a time evolved state given by
\begin{multline}
  \Lambda (t) \rho =  p_0 (t) \mathcal{F} (t) \rho +  \sum_{n = 1}^{\infty} \int^t_0 \mathd t_n \ldots
  \int^{t_2}_0 \mathd t_1  \label{eq:nmdyson}\\
 \times p_n ( t ; t_n, \ldots, t_1 )
  \mathcal{F} ( t - t_n ) \mathcal{E} 
\ldots \mathcal{E} \mathcal{F} ( t_1 ) \rho .
\end{multline}
Here $p_n ( t ; t_n, \ldots_, t_1 )$ denotes the exclusive
probability density for the realization of $n$ events up to time $t$, at given
times $t_1, \ldots, t_n$, with no events in between. This probability density
for a renewal process reads
\begin{equation}
  p_n ( t ; t_n, \ldots_, t_1 ) = f ( t - t_n )
  \ldots f ( t_2 - t_1 ) g ( t_1 ),  \label{eq:epd}
\end{equation}
with $f (t)$ a waiting time distribution, i.e. a distribution
function over the positive reals, and $g (t) = 1 - \int^t_0
\mathd \tau f (\tau)$ its associated survival probability,
expressing the probability that no jump has taken place up to time $t$. Thanks
to CPT of the maps $\mathcal{E}$ and
$\mathcal{F} (t)$ the obtained dynamics is indeed well defined.
CP is warranted by stability of the positive cone of CP maps
under composition. Regarding trace preservation, due to
Eq.~(\ref{eq:epd}) for a renewal process the probability $p_k (t)$
to have $k$ counts up to time $t$ obeys
\begin{equation}
  p_k (t) = \int^t_0 \mathd \tau f ( t - \tau )
  p_{k - 1} (\tau),  \label{eq:pk}
\end{equation}
with $p_0 (t) = g (t)$. Iterating this identity one
obtains $\tmop{Tr} \Lambda (t) \rho = \sum_{k = 0}^{\infty} p_k
(t) \rho = \rho$. The constructed collection of CPT time evolutions
$\Lambda (t)$ are functionals of $\mathcal{F} ( t
)$, $f (t) \mathcal{}$ and $\mathcal{E}$, and allows for
a simple operational interpretation in terms of the random action of a fixed
quantum channel over a given dynamics, not necessarily obeying a semigroup
composition law.

\paragraph*{Laplace transform and master equation.}

We now observe that, according to its definition Eq.~(\ref{eq:nmdyson}),
the map $\Lambda (t)$ obeys the integral equation
\begin{equation}
  \Lambda (t) = g (t) \mathcal{F} (t) + \int^t_0 \mathd \tau f ( t - \tau ) \mathcal{F}
  ( t - \tau ) \mathcal{E} \Lambda (\tau), 
  \label{eq:lt}
\end{equation}
which in Laplace transform, here denoted by a hat, simply reads
\begin{equation}
  \hat{\Lambda} (u) = \widehat{g \mathcal{F}} (u) + \widehat{f \mathcal{F}} (u) \mathcal{E} \hat{\Lambda}
  (u). \label{eq:lu}
\end{equation}
Starting from this expression, as described in the Supplemental
Material {\cite{sm}} one finally obtains the closed master equation
\begin{equation}
  \frac{\mathd}{\mathd t} \rho (t) = \int^t_0 \mathd \tau
  \mathcal{K} ( t - \tau ) \mathcal{E} \rho (\tau) +
  \mathcal{I} (t) \rho ( 0 ),  \label{eq:ms}
\end{equation}
with kernel and inhomogeneous term given by
\begin{equation}
\label{eq:k-s}
   \mathcal{K} (t)  \!=\!  \frac{\mathd}{\mathd t}[ f(t)\mathcal{F} (t) ] + f (0) \delta
   (t)  
\quad
   \mathcal{I} (t) \!=\! \frac{\mathd}{\mathd t} [ g(t) \mathcal{F} (t) ] .
\end{equation}
This is the main result of our Letter. We stress the fact that the map $\Lambda
(t)$, solution of Eq.~(\ref{eq:lt}), or equivalently
Eq.~(\ref{eq:ms}), is CPT by construction. It can be obtained as the
inverse Laplace transform of the solution of Eq.~(\ref{eq:lu})
\begin{equation}
  \hat{\Lambda} (u) = \left[ \mathbbm{1} - \widehat{f
  \mathcal{F}} (u) \mathcal{E} \right]^{- 1} \widehat{g
  \mathcal{F}} (u) .  \label{eq:sollu}
\end{equation}
This identity provides a compact general expression of the Laplace transform
of the exact solution, in terms of the transform of the elementary
maps determining the time evolution. Note that the result has been obtained
without making any restrictive assumption on the dimensionality of the Hilbert
space of the system.

\paragraph*{Limiting expressions}

Before considering the non-Markovianity (NM) of the
class of master equations introduced above in view of the recently
proposed criteria \cite{Breuer2009b,Rivas2010a},
we to point to some special cases already considered in
the literature.
Firstly a quantum dynamical semigroup is recovered if
$\mathcal{F} (t) \rightarrow \mathe^{t \mathcal{L}}$, with
$\mathcal{L}$ in Lindblad form, and $\mathcal{E} \rightarrow
\mathbbm{1}$, independently of the waiting time distribution $f
(t)$. Indeed, the solution given by
Eq.~(\ref{eq:sollu}) thanks to the properties of the Laplace transform with
respect to shifts now reads $\hat{\Lambda} (u) = \sum^{\infty}_{k =
0} \hat{g} ( u - \mathcal{L} ) \hat{f}^k ( u -
\mathcal{L} )$, and therefore, also using $\hat{p}_k (u) =
\hat{g} (u) \hat{f}^k (u)$, which
follows from Eq.~(\ref{eq:pk}), we have $\rho (t) = \mathe^{t
\mathcal{L}} \rho$. More generally, for
a non trivial CPT map $\mathcal{E}$ rearranging terms one
obtains {\cite{sm}}
\begin{equation}
  \frac{\mathd}{\mathd t} \rho (t)  =  \mathcal{L} \rho (t) +\int^t_0  \mathd \tau k ( t - \tau )
  \mathe^{( t - \tau ) \mathcal{L}} \mathcal{\left[ E - \mathbbm{1}
  \right]} \rho (\tau),  \label{eq:budini}
\end{equation}
where the $\mathbbm{C}$-number kernel reads $\hat{k} (u) =
\hat{f} (u) / \hat{g} (u)$. This equation has been
previously considered for the special case of a Lindblad generator given by a
simple commutator, pointing to a possible
microscopic derivation {\cite{Budini2004a,Budini2005a}}. For a
vanishing Lindblad generator one has in particular $\rho (t) =
\sum^{\infty}_{k = 0} p_k (t) \mathcal{E}^k \rho$, a class of
non-Markovian evolutions studied in
{\cite{Budini2004a,Vacchini2011a,Vacchini2012a}}.

If we allow for a generic CPT map $\mathcal{F} (t)$, but do
consider the events as a reset of the continuous time dynamics
described by $\mathcal{F} (t)$, so that $\mathcal{E} \rightarrow
\mathbbm{1}$, we end up with
\begin{equation}
  \frac{\mathd}{\mathd t} \rho (t) = \int^t_0  \mathd \tau f
  ( t - \tau ) \mathcal{F} ( t - \tau ) \dot{\rho}
  (\tau) + g (t) \dot{\mathcal{F}} (t) \rho,  \label{eq:giovannetti}
\end{equation}
which for the case of a memoryless waiting time of exponential type, $f (
t ) = \Gamma \mathe^{- \Gamma t}$, so that $g (t) =
\mathe^{- \Gamma t}$, recovers the result recently obtained relying on a collisional
model assuming collisions with independent ancillas {\cite{Ciccarello-xxx}}.

\paragraph*{Non-Markovianity.\label{sec:nmy}}

We now study the NM of the dynamics described by the master equation
Eq.~(\ref{eq:ms}). Indeed, despite the fact that the considered master
equation can include more general situations than a semigroup dynamics
generated by a Lindblad operator, the degree of NM of the obtained dynamics is
still to be ascertained. To this aim we will make reference to the definition
of NM associated to the idea of revival of distinguishability among different
states advocated in {\cite{Breuer2009b,Breuer2012a}}, considering the trace
distance as a natural quantifier of distinguishability. As it has
been shown, this criterion is more stringent than the violation of
divisibility in terms of CP maps
{\cite{Laine2010a,Mazzola2010a,Haikka2011a,Vacchini2011a}}. As a result, if we
detect NM by using the notion of distinguishability, we know that the
considered dynamics is non-Markovian also from the divisibility point of view.
We recall that the trace distance between two states $\rho_1 (t)$
and $\rho_2 (t)$ is given by the trace norm of their difference
$D ( \rho_1 (t), \rho_2 (t) ) =
\frac{1}{2} \| \rho_1 (t) - \rho_2 (t) \|_1$, that
is the sum of the modulus of the eigenvalues of their difference. It takes
values between zero and one and can be interpreted as a measure of the
distinguishability among states. In particular, relying on the fact that the
trace distance is a contraction with respect to the action of a CPT map, M of
the map is identified with the monotonic decrease in time of the trace
distance among any couple of possible initial states. NM is then detected
whenever the time derivative of the trace distance grows at a certain time
$t$, for at least a couple of initial states, i.e. $\dot{D} ( \rho_1
(t), \rho_2 (t) ) > 0$. In order to highlight
this behavior, let us make specific choices for the system and the different
maps and functions determining the time evolution $\Lambda (t)$.
We therefore consider the Hilbert space $\mathbbm{C}^2$, and take as CPT map
$\mathcal{E}$ a Pauli channel $\mathcal{E}_i \rho = \sigma_i \rho \sigma_i$,
with $i = 0, x, y, z$ and $\sigma_0 = \mathbbm{1_{}}$. We further take as
waiting time distribution $f (t)$ a convolution of exponential
distributions. These waiting time
distributions bring with themselves a natural time scale given by the mean
waiting time. Finally we have the freedom to consider a collection of time
dependent CPT maps. The latter also have an intrinsic time scale, and the
interplay between the two time scales plays an important role in the
characterization of NM. To this aim we will analyze two situations,
corresponding to different physical implementations. As a first example we
take a map $\mathcal{F}_d (t)$ only affecting coherences, which
according to the trace distance criterion by itself always describes a
non-Markovian dynamics. As a complementary situation we will deal with a time
evolution $\mathcal{F}_+ (t)$ which itself admits both a
Markovian and a non-Markovian limit, and affects all components of the
statistical operator.

\paragraph{Examples.}

\begin{figure}[tb]
  \resizebox{40mm}{40mm}{\includegraphics{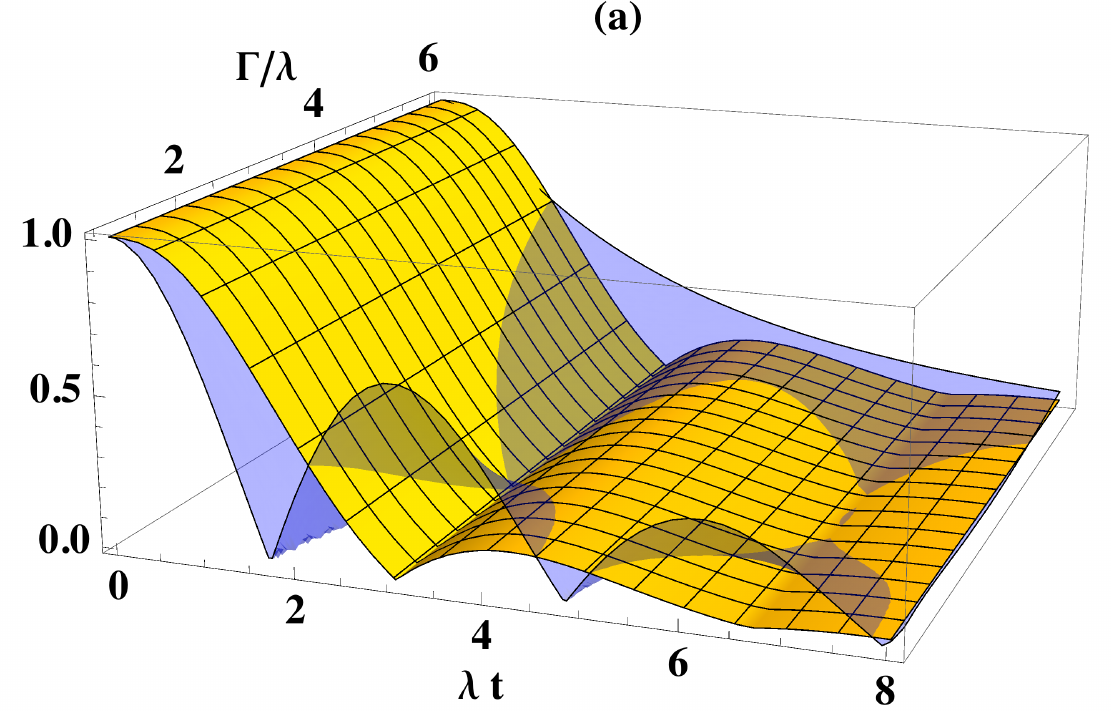}}\resizebox{40mm}{40mm}{\includegraphics{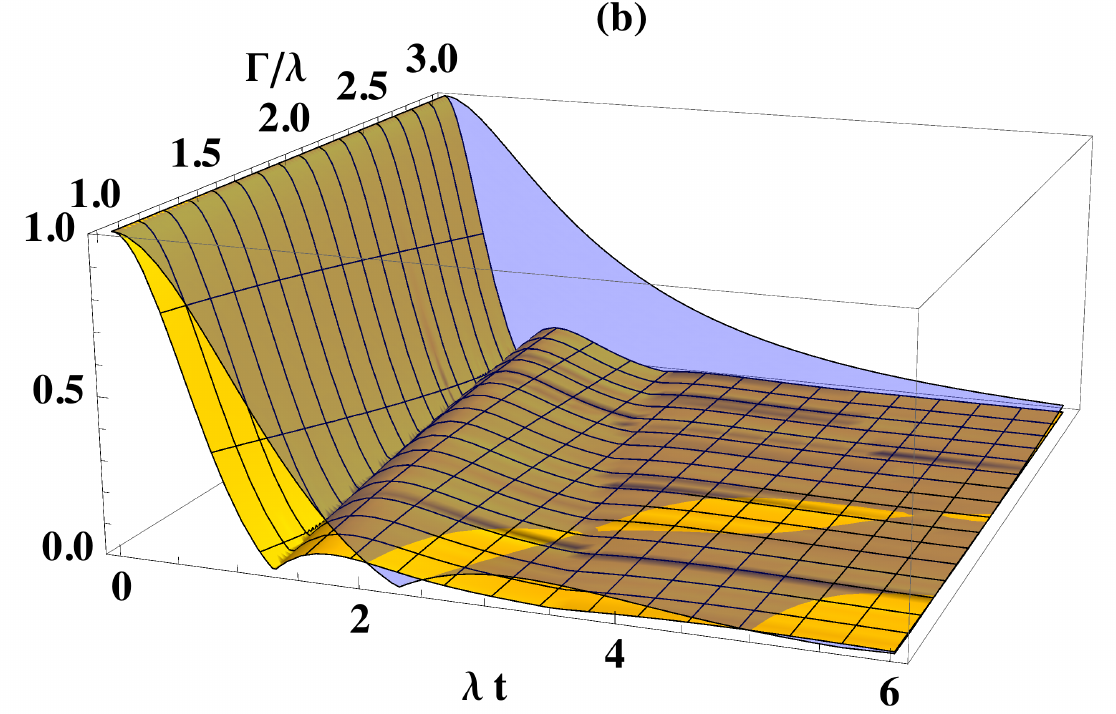}}
  \caption{\label{fig:nm-plot}(Color online) (a) Modulus of the
  functions $d_- (t)$ and $q (t)$ for
  a dephasing dynamics described by $D (t) = \cos ( \lambda
  t )$, and waiting time given by the convolution of
  three equal exponentials $\Gamma \mathe^{- \Gamma t}$. The growth of any of
  these quantities, as discussed in the Supplemental Material {\cite{sm}},
  provides a direct signature of NM of the time evolution map $\Lambda_d
  (t)$. The quantities are plotted as a function of $\lambda t$
  and $\Gamma / \lambda$. The semitransparent surface corresponds to $d_-
  (t)$, while the meshed surface represents $q (t)$.
  It immediately appears that for growing ratio $\Gamma / \lambda$,
  determining the relation between the time scales inherent in $\mathcal{F}_d
  (t)$ and $f (t)$, the oscillations in $d_- (t)$ are suppressed. The NM is then only detected by $q ( t
  )$, arising due to the action of the map $\mathcal{E}_x$ which
  describes the events, in this case spin flips, in between the continuous
  time evolution $\mathcal{F}_d (t)$. (b) Modulus of $g_- (t)$ and $h_+ (t)$, here for a
  continuous dynamics $\mathcal{F}_+ (t)$ involving both
  populations and coherences, and waiting time corresponding to the
  convolution of two equal exponential distributions. The ratio $\gamma / \lambda$
  appearing in the function $G (t)$ given by Eq.~(\ref{eq:G}) is
  set equal to 3, corresponding to NM of $\mathcal{F}_+ (t)$
  alone. Again the growth of the modulus of any of these functions warrants
  NM. The function $g_- (t)$ given by the semitransparent surface
  only detects NM for small $\Gamma$, while $h_+ (t)$ shows that
  NM also takes place for frequent events, that is large $\Gamma / \lambda$,
  even though it is confined to shorter and shorter times.}
\end{figure}

We first consider a dephasing dynamics $\mathcal{F}_d (t)$, which
multiplies the off-diagonal matrix elements of the statistical operator by
the function $D (t)$. Working in $\mathbbm{C}^2$ it is convenient
to represent statistical operators through their coefficients on the linear
basis $\left\{ \sigma_i / \sqrt{2} \right\}$, so that maps can be represented
as matrices {\cite{Andersson2007a,Smirne2010b}}. This dephasing map in
particular is represented as a diagonal matrix $F_d (t) =
\tmop{diag} ( 1, D (t), D (t), 1 )$, and
the same holds for the Pauli maps which take the general form $E = \tmop{diag}
( 1, \varepsilon_x, \varepsilon_y, \varepsilon_z )$, with
$\varepsilon_i = \pm 1$, the sign depending on the specific choice of map.
Relying on Eq.~(\ref{eq:sollu}), these expressions after some algebra
{\cite{sm}} lead to the following compact result for the time evolution map
\begin{equation}
  \Lambda_d (t) = \tmop{diag} ( 1, X (t), Y (t), Z (t) ) .  \label{eq:ld}
\end{equation}
For the expression of the time dependent functions appearing in the
evolution map we consider the functional
\begin{equation}
  \hat{L}^{\pm}_f \left[ M \right] (u) = \frac{\widehat{gM}
  (u)}{1 \pm \widehat{fM} (u)}, 
  \label{eq:functional}
\end{equation}
where $M$ denotes an arbitrary function of time. $X (t)$ and $Y
(t)$ are then given by one of the functions $d_{\pm} ( t
) = L^{\pm}_f \left[ D \right] (t)$, depending on the value
of the $\varepsilon_i$, while $Z (t)$ is given by either the
identity or the function $q (t) = \sum^{\infty}_{n = 0} p_{2 n}
(t) - \sum^{\infty}_{n = 0} p_{2 n + 1} (t)$, which gives the difference
between the probability to have an even and an odd number of jumps. Given the
explicit expression of the map, one can calculate the time
derivative of the trace distance among two different initial states, which
shows in particular that one has NM whenever the modulus of one of the
functions $d_{\pm} (t)$ or $q (t)$ grows, as
discussed in the Supplemental Material {\cite{sm}}. This case is depicted in
Fig.~\ref{fig:nm-plot}(a), considering a dephasing map $D (t) = \cos (
\lambda t )$. In this case the dynamics given by $\mathcal{F}_d ( t
)$ alone never allows for a Markovian description. Here the rate
$\lambda$ sets the natural time scale for this contribution to the dynamics,
to be compared with the time scale $1 / \Gamma$ given by the mean waiting time
associated to the waiting time distribution $f (t)$. As it
appears in Fig.~\ref{fig:nm-plot}(a), if $\Gamma / \lambda \gg 1$, so that
subsequent events are very close in time, the contribution to NM due to
$\mathcal{F}_d (t)$ is suppressed, since on a short enough time
any time evolution map is Markovian.

As a further example we consider the dynamical map $\mathcal{F}_+ ( t
)$, affecting both populations and coherences, that arises considering
the interaction of a two-level system with a bosonic field in the
vacuum state {\cite{Breuer2007}}. The map is characterized by the
function $G (t)$, depending on the spectral density of the
environment, and in matrix form reads
\begin{equation}
  F_+ (t)\!=\! \tmop{diag} ( 1, G (t), G
  (t), \left| G (t) \right|^2 ) + B (
  \left| G (t) \right|^2 \!-\! 1 ),  \label{eq:fpiu}
\end{equation}
where $B ( x )$ denotes the $4 \times 4$ matrix with entry $x$ in
the bottom left corner as the only non zero element.
Exploiting Eq.~(\ref{eq:sollu}) we can obtain the expression of the time
evolution map {\cite{sm}}
\begin{equation}
  \Lambda_+ (t) = \tmop{diag} ( 1, X (t), Y
  (t), Z (t) ) + B ( W (t)
  ),  \label{eq:lpiu}
\end{equation}
where now $X (t)$ and $Y (t)$ take the expressions
$g_{\pm} (t) = L^{\pm}_f \left[ G \right] (t)$,
while $Z (t)$ corresponds to $h_{\pm} (t) =
L^{\pm}_f [ | G |^2 ] (t)$. The function $W (
t )$, determined by $f (t)$ and $| G ( t
) |^2$, does not affect the trace distance, since it corresponds to a fixed translation of the state
{\cite{Wissmann2012a}}. A typical expression of $G (t)$ is given
by
\begin{equation}
  G (t) = \mathe^{- \lambda t / 2} \left[ \cosh (
  {\tilde{\gamma} t}/{2}) + ({\lambda}/{\tilde{\gamma}})\sinh
  ( {\tilde{\gamma} t}/{2}) \right],  \label{eq:G}
\end{equation}
where $\tilde{\gamma} = \sqrt{\lambda^2 - 2 \gamma \lambda}$, and has the
interesting feature that for $\gamma / \lambda < 1 / 2$ the map $\mathcal{F}_+
(t)$ itself is Markovian, while for $\gamma /
\lambda$ above this threshold one has NM {\cite{Laine2010a}}. The NM of the
ensuing overall dynamics $\Lambda (t)$ is considered in
Fig.~\ref{fig:nm-plot}(b), where we have plotted the modulus of the functions
$g_{-} (t)$ and $h_{+} (t)$ for $G(t)$ as in Eq.~(\ref{eq:G}). Again the growth of the
modulus of any of these functions is a witness of NM. It appears indeed that
for a wide range of parameters the dynamics is non-Markovian, yet the NM is
actually the result of an interplay of the features of all the three elements
determining the dynamics, namely $\mathcal{F}_+ (t)$,
$\mathcal{E}$ and $f (t)$. Indeed for values of the ratio $\gamma
/ \lambda$ such that $\mathcal{F}_+ (t)$ itself is non-Markovian,
the dynamics $\Lambda (t)$ might still be Markovian, if the ratio
$\Gamma / \lambda$ of the time scales associated to $\mathcal{F}_+ ( t
)$ and $f (t)$ is high enough. On the contrary, even a
Markovian $\mathcal{F}_+ (t)$ can give rise to a non-Markovian
dynamics because of the action of the map $\mathcal{E}$ in between the
continuous time evolutions, and of the distribution in time of these events.

\paragraph*{Conclusions.}

We have obtained a large set of closed non-Markovian master equations
starting from a piecewise dynamics described by a continuous time
evolution interrupted by random jumps. The solution of these equations is warranted to be a CPT map. These
master equations involve both a memory kernel and a inhomogeneous
term. The basic ingredients in the construction are a collection of
time dependent maps, together with a waiting time distribution
describing the random occurrence of events characterized by a quantum
channel. We have considered the connection of this result with the
standard expression of quantum dynamical semigroups, as well as more
recent examples of non-Markovian master equations obtained starting
from microscopic models. In particular, we have certified the NM of
the obtained time evolution by studying the behavior in time of the
distinguishability between two different initial states, as quantified
by the trace distance. Finally, the operational interpretation of the
structure of these master equations paves the way for their use in
concrete applications.

\paragraph*{Acknowledgments.}

The author thanks A. Smirne for discussions and
reading of the manuscript. Support from COST Action MP 1006 is gratefully
acknowledged.

\section*{Supplemental material}

In this Supplemental material we provide technical details on the derivation
of equations and properties discussed in the main text of the paper.

\subsubsection*{Derivation of Eq.~(\ref{eq:ms})}

We here derive the closed master equation obeyed by the statistical operator
$\rho (t)$. Given that $\Lambda (t)$ obeys the
integral equation Eq.~(\ref{eq:lt}), as considered in the main text in Laplace
transform the equation for $\hat{\Lambda} ( u )$ reads
\begin{eqnarray}
  \hat{\Lambda} ( u ) & = & \widehat{g \mathcal{F}} ( u
  ) + \widehat{f \mathcal{F}} ( u ) \mathcal{E} \hat{\Lambda}
  ( u ), \nonumber
\end{eqnarray}
with the Laplace transform defined as usual and denoted by a hat $\hat{h} ( u )
= \int^{\infty}_0 \mathd t \mathe^{- ut} h (t)$,
so that multiplying by $u$ and subtracting the identity operator from both
sides, at the same adding and subtracting the term $f ( 0 )
\mathcal{E} \hat{\Lambda} ( u )$ at the l.h.s. one comes to
\begin{eqnarray}
  u \hat{\Lambda} ( u ) - \mathbbm{1} & = & \left[ u \widehat{g
  \mathcal{F}} ( u ) - \mathbbm{1} \right] + \left[ u \widehat{f
  \mathcal{F}} ( u ) - f ( 0 ) \right] \mathcal{E}
  \hat{\Lambda} ( u ) \nonumber\\
  &  & + f ( 0 ) \mathcal{E} \hat{\Lambda} ( u ),
  \nonumber
\end{eqnarray}
so that recalling that the Laplace transform of the derivative of a function
$h (t)$ is given by $u \hat{h} ( u ) - \mathbbm{1}$,
and using $\mathcal{F} ( 0 ) = \mathbbm{1}$, one obtains
\begin{eqnarray}
  \frac{\mathd}{\mathd t} \Lambda (t) & = & \int^t_0 \mathd \tau
  \frac{\mathd}{\mathd ( t - \tau )} f ( t - \tau )
  \mathcal{F} ( t - \tau ) \mathcal{E} \Lambda ( \tau )
  \nonumber\\
  &  & + f ( 0 ) \mathcal{E} \Lambda (t) + \frac{\mathd}{\mathd t} \left[ g (t) \mathcal{F} (
  t ) \right] . \nonumber
\end{eqnarray}
According to the relation $\rho (t) = \Lambda (t)
\rho ( 0 )$ and using the identifications Eq.~(\ref{eq:k-s}) one
finally comes to the master equation Eq.~(\ref{eq:ms}). \

\subsubsection*{Derivation of Eq.~(\ref{eq:budini}) and
Eq.~(\ref{eq:giovannetti})}

In order to derive the master equation Eq.~(\ref{eq:budini}) we start from
Eq.~(\ref{eq:lu}) and take $\mathcal{F} (t)$ to be of exponential
form $\mathe^{t \mathcal{L}}$ with $\mathcal{L}$ a Lindblad generator. Thanks
to the behavior of the Laplace transform with respect to shifts one thus has
$\widehat{g \mathcal{F}} ( u ) = \widehat{g \mathcal{}} ( u -
\mathcal{L} )$, and similarly for $\widehat{f \mathcal{F}} ( u
)$, so that for the Laplace transform of the statistical operator
$\hat{\rho} ( u )$ we obtain
\begin{eqnarray}
  \hat{\rho} ( u ) & = & \widehat{g \mathcal{}} ( u -
  \mathcal{L} ) \rho ( 0 ) + \widehat{f \mathcal{}} ( u
  - \mathcal{L} ) \mathcal{E} \tilde{\rho} ( u ) . 
  \label{eq:k1}
\end{eqnarray}
To proceed further we note that from the relation between waiting time
distribution and survival probability
\begin{eqnarray}
  g (t) & = & 1 - \int^t_0 \mathd \tau f ( \tau ), 
  \label{eq:gf}
\end{eqnarray}
one has
\begin{eqnarray}
  \hat{g} ( u ) & = & \frac{1 - \hat{f} ( u )}{u},
  \nonumber
\end{eqnarray}
and therefore introducing the function
\begin{eqnarray}
  \hat{k} ( u ) & = & \frac{\hat{f} ( u )}{\hat{g}
  ( u )},  \label{eq:k2}
\end{eqnarray}
also
\begin{eqnarray}
  \frac{1}{\widehat{g \mathcal{}} ( u - \mathcal{L} )} - \widehat{k
  \mathcal{}} ( u - \mathcal{L} ) & = & u -\mathcal{L}. \nonumber
\end{eqnarray}
We note that the function $k (t)$ naturally appears as memory
kernel in the description of continuos time random walks {\cite{Hughes1995}}.
Dividing Eq.~(\ref{eq:k1}) by $\widehat{g \mathcal{}} ( u - \mathcal{L}
)$ and using Eq.~(\ref{eq:k2}) one thus obtains, subtracting a term
$\widehat{k \mathcal{}} ( u - \mathcal{L} ) \hat{\rho} ( u
)$ from both sides
\begin{eqnarray}
  u \hat{\rho} ( u ) - \rho ( 0 ) & = & \mathcal{L}
  \hat{\rho} ( u ) + \widehat{k \mathcal{}} ( u - \mathcal{L}
  ) \left[ \mathcal{E}- \mathbbm{1} \right] \tilde{\rho} ( u
  ),  \label{eq:k3}
\end{eqnarray}
and finally taking the inverse Laplace transform, exploiting again the
property of the Laplace transform with respect to shifts, the master equation
\begin{eqnarray}
  \frac{\mathd}{\mathd t} \rho (t) & = & \mathcal{L} \rho (
  t ) + \int^t_0 \mathd \tau k ( t - \tau ) \mathe^{( t
  - \tau ) \mathcal{L}} \mathcal{\left[ E - \mathbbm{1} \right]} \rho
  ( \tau ) . \nonumber
\end{eqnarray}
For the case of Eq.~(\ref{eq:giovannetti}) we start from Eq.~(\ref{eq:lu})
and again multiply by $u$ and subtract the identity from both sides, so that
suitably rearranging terms and taking $\mathcal{E} \rightarrow \mathbbm{1}$ we
have
\begin{eqnarray}
  u \hat{\Lambda} ( u ) - \mathbbm{1} & = & \left[ u \widehat{g
  \mathcal{F}} ( u ) - \mathbbm{1} \right] + \widehat{f
  \mathcal{F}} ( u ) \left[ u \hat{\Lambda} ( u ) -
  \mathbbm{\mathbbm{1}} \right] \nonumber\\
  &  & + \widehat{f \mathcal{F}} ( u ), \nonumber
\end{eqnarray}
leading to
\begin{eqnarray}
  \frac{\mathd}{\mathd t} \Lambda (t) & = & \int^t_0 \mathd \tau
  f ( t - \tau ) \mathcal{F} ( t - \tau ) \dot{\Lambda}
  ( \tau ) + f (t) \mathcal{F} (t)
  \nonumber\\
  &  & + \frac{\mathd}{\mathd t} \left[ g (t) \mathcal{F} (
  t ) \right], \nonumber
\end{eqnarray}
and further exploiting the relation $\dot{g} (t) = - f ( t
)$ following from Eq.~(\ref{eq:gf}) one finally obtains the desired
master equation Eq.~(\ref{eq:giovannetti})
\begin{eqnarray}
  \frac{\mathd}{\mathd t} \Lambda (t) & = & \int^t_0 \mathd \tau
  f ( t - \tau ) \mathcal{F} ( t - \tau ) \dot{\rho}
  ( \tau ) + g (t) \mathcal{\dot{F}} (t)
  . \nonumber
\end{eqnarray}

\subsubsection*{Derivation of the map $\Lambda_d (t)$}

We now derive the time evolution map $\Lambda_d (t)$ for a
dephasing dynamics, which only affects the off-diagonal matrix elements of the
statistical operator of the system, multiplying them by a function $D ( t
)$, taken in the example to be $\cos ( \lambda t )$. Any
statistical operator on $\mathbbm{C}^2$ can be represented by a vector in the
linear basis $\left\{ \frac{1}{\sqrt{2}} \mathbbm{1}, \frac{1}{\sqrt{2}}
\sigma_x, \frac{1}{\sqrt{2}} \sigma_y, \frac{1}{\sqrt{2}} \sigma_z \right\}$,
orthonormal according to the Hilbert-Schmidt scalar product, so that maps can
be identified with suitable $4 \times 4$ matrices \
{\cite{Andersson2007a,Smirne2010b}}. The dephasing map $\mathcal{F}_d ( t
)$ in this basis acts as the diagonal matrix
\begin{eqnarray}
  F_d (t) & = & \tmop{diag} ( 1, D (t), D
  (t), 1 ), \nonumber
\end{eqnarray}
while the Pauli map $\mathcal{E}_x$ corresponds to the diagonal matrix
\begin{eqnarray}
  E_x & = & \tmop{diag} ( 1, 1, - 1, - 1 ) . \nonumber
\end{eqnarray}
Starting from this result we have that the Laplace transform of the operator
$f (t) \mathcal{F}_d (t)$ can be written as
$\tmop{diag} ( \hat{f} ( u ), \widehat{fD} ( u ),
\widehat{fD} ( u ), \hat{f} ( u ) )$, and
similarly for $g (t) \mathcal{F}_d (t)$. Thanks to
the closure of the algebra of diagonal matrices $\hat{\Lambda}_d ( u
)$ itself turns out to be diagonal, and according to
Eq.~(\ref{eq:sollu}) reads
\begin{eqnarray}
  \hat{\Lambda}_d ( u ) & = & \tmop{diag} \left( \frac{1}{u},
  \frac{\widehat{fD} ( u )}{1 - \widehat{fD} ( u )},
  \frac{\widehat{fD} ( u )}{1 + \widehat{fD} ( u )},
  \frac{1}{u} \frac{1 - \hat{f} ( u )}{1 + \hat{f} ( u
  )} \right) . \nonumber
\end{eqnarray}
Upon defining
\begin{eqnarray}
  d_{\pm} (t) & = & L^{\pm}_f \left[ D \right] (t) =
  \frac{\widehat{gD} ( u )}{1 \pm \widehat{fD} ( u )}
  \nonumber
\end{eqnarray}
as in Eq.~(\ref{eq:functional}), as well as
\begin{eqnarray}
  \hat{q} ( u ) & = & \frac{1}{u} \frac{1 - \hat{f} ( u
  )}{1 + \hat{f} ( u )}, \nonumber
\end{eqnarray}
which according to the relation $\hat{p}_k ( u ) = \widehat{g
\mathcal{}} ( u ) \hat{f}^k ( u )$, which follows from
Eq.~(\ref{eq:pk}), is the Laplace transform of the quantity $q ( t
) = \sum^{\infty}_{n = 0} p_{2 n} (t) - \sum^{\infty}_{n = 0} p_{2 n +
1} (t)$, we finally obtain
\begin{eqnarray}
  \Lambda_d (t) & = & \tmop{diag} ( 1, d_- (t),
  d_+ (t), q (t) ), \nonumber
\end{eqnarray}
which provides the explicit expression of Eq.~(\ref{eq:ld}) when the Pauli
channel is given by $\mathcal{E}_x$. Similar results apply for the other Pauli
channels. The modulus of the functions $d_- (t)$ and $q (
t )$ is plotted in Fig.~\ref{fig:nm-plot}(a), since it provides evidence
for NM of the dynamics, as discussed in the next paragraph.

\subsubsection*{Non-Markovianity of the time evolution map}

We here apply the trace distance criterion for the detection of NM to the
dynamics described by the map $\Lambda_d (t)$, and similar
conclusions hold for $\Lambda_+ (t)$. As discussed in the main
text, according to this criterion NM is associated to the growth of the
distinguishability in time, as quantified by the trace distance, of two
distinct initial states. Given two initial states $\rho_1 ( 0 )$
and $\rho_2 ( 0 )$ one monitors their trace distance in time, as
given by
\begin{eqnarray}
  D ( \rho_1 (t), \rho_2 (t) ) & = &
  \frac{1}{2} \| \rho_1 (t) - \rho_2 (t) \|_1
  \nonumber\\
  & = & \frac{1}{2} \| \Lambda_d (t) ( \rho_1 ( 0
  ) - \rho_2 ( 0 ) ) \|_1, \nonumber
\end{eqnarray}
and the map is said to be non-Markovian if there exist a couple of initial
states and a point in time such that their distinguishability grows, i.e.
\begin{eqnarray}
  \frac{\mathd}{\mathd t} D ( \rho_1 (t), \rho_2 ( t
  ) ) & > & 0. \nonumber
\end{eqnarray}
For the case at hand, setting $\Delta_p$ for the difference in the populations
of the two initial statistical operators, as well as $\Delta_c$ for the
difference in the coherences, that is the off-diagonal matrix element, for a
map diagonal in the basis used to represent states as vectors the trace
distance and its derivative can be explicitly calculated. Using as in
Eq.~(\ref{eq:ld}) the notation
\begin{eqnarray}
  \Lambda_d (t) & = & \tmop{diag} ( 1, X (t), Y
  (t), Z (t) ) \nonumber
\end{eqnarray}
we have
\begin{multline}
  D ( \rho_1 (t), \rho_2 (t) ) =
  \nonumber\\
 \sqrt{\Delta^2_p Z^2 (t) + \tmop{Re}^2 \Delta_c X^2
  (t) + \tmop{Im}^2 \Delta_c Y^2 (t)}, \nonumber
\end{multline}
and therefore
\begin{multline}
  \frac{\mathd}{\mathd t} D ( \rho_1 (t), \rho_2 ( t
  ) ) =  \nonumber\\
   \frac{1}{2} \frac{\Delta^2_p \frac{\mathd}{\mathd t} Z^2 ( t
  ) + \tmop{Re}^2 \Delta_c \frac{\mathd}{\mathd t} X^2 (t)
  + \tmop{Im}^2 \Delta_c \frac{\mathd}{\mathd t} Y^2 ( t
  )}{\sqrt{\Delta^2_p Z^2 (t) + \tmop{Re}^2 \Delta_c X^2
  (t) + \tmop{Im}^2 \Delta_c Y^2 (t)}}, \nonumber
\end{multline}
so that one has growth of the trace distance if the modulus of any of the
functions $X (t)$, $Y (t)$ or $Z (t)$
grows.

\subsubsection*{Derivation of the map $\Lambda_+ (t)$}

We now consider as Pauli map $\mathcal{E}_z$, and introduce a continuous time
dynamics determined by a map $\mathcal{F}_+ (t)$ which in the
above introduced basis for the operators in $\mathbbm{C}^2$ is expressed as in
Eq.~(\ref{eq:fpiu}) by the matrix
\begin{eqnarray}
  F_+ (t) & = & \tmop{diag} ( 1, G (t), G
  (t), \left| G (t) \right|^2 ) + B (
  \left| G (t) \right|^2 - 1 ), \nonumber
\end{eqnarray}
where as discussed in the main text the matrix $B ( x )$ has the
only non zero entry $x$ in the bottom left corner. The calculations closely
follow those performed for $\Lambda_d (t)$. In particular thanks
to the closure of the algebra of matrices with non zero entries only on the
diagonal and in the bottom left corner, which are such that the inverse if it
exists still is in the algebra, relying on Eq.~(\ref{eq:sollu}) we obtain
\begin{eqnarray}
  \hat{\Lambda}_+ ( u ) & = & \tmop{diag} \left( \frac{1}{u},
  \frac{\widehat{fG} ( u )}{1 + \widehat{fG} ( u )},
  \frac{\widehat{fG} ( u )}{1 + \widehat{fG} ( u )},
  \frac{\widehat{f \left| G \right|^2} ( u )}{1 - \widehat{f \left|
  G \right|^2} ( u )} \right) \nonumber\\
  &  & + B ( \hat{W} ( u ) ), \nonumber
\end{eqnarray}
where
\begin{eqnarray}
  \hat{W} ( u ) & = & \frac{1}{u} \frac{2 \hat{f} ( u )
  - 1 + u \left( \widehat{g \left| G \right|^2} ( u ) - \widehat{f
  \left| G \right|^2} ( u ) \right)}{1 + \widehat{f \left| G
  \right|^2} ( u )} . \nonumber
\end{eqnarray}
According to the definitions given in the main text below Eq.~(\ref{eq:lpiu})
we finally arrive to
\begin{eqnarray}
  \Lambda_+ (t) & = & \tmop{diag} ( 1, h_+ (t),
  h_+ (t), g_- (t) ) + B ( W ( t
  ) ) . \nonumber
\end{eqnarray}
Considering a map of the form
\begin{eqnarray}
  \Lambda_+ (t) & = & \tmop{diag} ( 1, X (t), Y
  (t), Z (t) ) + B ( W (t)
  ) \nonumber
\end{eqnarray}
one immediately sees that the term $W (t)$ provides a
contribution to the matrix elements of the statistical operator which is
independent of the initial state, so that it does not affect the behavior of
the trace distance. As a result also in this case the M or NM of the map
$\Lambda_+ (t)$ does depend on the behavior of the modulus of the
time dependent functions appearing on the diagonal of the matrix
representation of $\Lambda_+ (t)$. The latter is plotted in
Fig.~\ref{fig:nm-plot}(b).

\end{document}